\documentclass [12pt,a4paper]{article}
\usepackage{amssymb}
\newcommand{\bv}{\nobreak\hfill $\square$}

\newcommand{\proof}{\noindent{\it Proof. }}

\def\bbbr{{\mathbb R}}
\def\bbbc{{\mathbb C}}

\def\Tr{\mbox{Tr}\,}

\def\Diag{\mbox{Diag}\,}
\def\tr{\mbox{Tr}\,}

\def\im{\text{i}}

\def\aa{\alpha}

\def\iM{{\cal M}}

\def\iH{{\cal H}}
\def\iA{{\cal A}}
\def\A{{\cal D}_1}
\def\iD{{\cal D}}

\def\iK{{\cal K}}

\def\BH{B({\cal H})}
\def\BK{B({\cal K})}
\def\<{\langle}
\def\>{\rangle}
\def\bz{\left(}
\def\jz{\right)}
\def\M{{\mathcal M}}
\newcommand{\s}{\mbox{ }}

\def\H{{\mathcal H}}

\def\H{{\mathcal H}}
\def\A{{\mathcal A}}

\def\B{{\mathcal B}}

\def\Ad{\mbox{Ad}}
\def\s{\mbox{ }}
\def\ds{\s\s}

\def\P(m,j){P_{m,d}}
\usepackage{array}
\usepackage{amsmath} 
\newtheorem{thm}{Theorem}
\newtheorem{lemma}{Lemma}
\newtheorem{proposition}{Proposition}

\begin{document}
\ \vskip 1cm
\centerline{\LARGE {\bf Structure of sufficient}}
\medskip
\centerline{\LARGE {\bf quantum coarse-grainings}\footnote{
The work was supported by the Hungarian OTKA T032662.}}
\bigskip
\bigskip
\medskip
\centerline{\large Mil\'an Mosonyi\footnote{E-mail: 
mosonyi{@}math.bme.hu} and D\'enes Petz\footnote{E-mail: 
petz{@}math.bme.hu}}
\bigskip

\bigskip
\centerline{Department for Mathematical Analysis}
\centerline{Budapest University of Technology and Economics}
\centerline{ H-1521 Budapest XI., Hungary}
\bigskip\bigskip
\noindent
\begin{quote}
{\bf Abstract:} Let $\iH$ and $\iK$ be finite dimensional Hilbert spaces, 
$T : \BH \to \BK$ 
be a coarse-graining and $D_1$, $D_2$ be density matrices on $\iH$. In this
paper the consequences of the existence of a coarse-graining $\beta: 
\BK \to \BH$ satisfying $\beta T(D_s)=D_s$  are given. (This means that
$T$ is sufficient for $D_1$ and $D_2$.) It
is shown that $D_s=\sum_{p=1}^r \lambda_s(p) S_s^\iH(p)R^\iH(p)$ $(s=1,2)$
should hold with pairwise orthogonal summands and with commuting factors and
with some probability distributions $\lambda_s(p)$ for $1 \leq p \leq r$ 
$(s=1,2)$. This decomposition allows to deduce the exact condition for 
equality in the strong subaddivity of the von Neumann entropy.
\end{quote}

\begin{quote}
\end{quote}

\begin{quote}
\end{quote}

\section{Introduction}
Let $\iH$ and $\iK$ be finite dimensional Hilbert spaces and
let $T : \BH \to \BK$ be a trace-preserving completely positive 
(or at least 2-positive) mapping. The mapping $T$ sends density
matrices acting on $\iH$ into density matrices acting on $\iK$. 
Such a mapping is called channeling transformation in quantum
information theory, if $\iH= \iK$, then $T$ may describe the 
dynamical change of state. We use the term coarse-graining, because the
statistical aspects get emphasis. Let $D_1$ be a density of a quantum
state  on $\iH$. Then the coarse-grained density $T(D_1)$ contains 
less information about the original quantum state and provides a partial
knowledge of $D_1$. The statistical inference is manifested by a mapping
$\beta : \BK \to \BH$ and in the good case $\beta T (D_1)=D_1$ and the 
original state is recovered.

In this paper, we study the scenario, where two states, density matrices 
$D_1$ and $D_2$, are given and we want to distinguish between them. If 
this is not more difficult than distinguishing between $T(D_1)$ and 
$T(D_2)$, then the coarse-graining is called sufficient for this pair. 
Formally we say that $T$ is sufficient for $D_1$ and $D_2$ if there exists 
a trace preserving 2-positive mapping $\beta : \BK \to \BH$ such that
\begin{equation}\label{E:M}
\beta T (D_1)=D_1 \quad \mbox{and}\quad \beta T (D_2)=D_2.
\end{equation}
This $\beta$ plays the role of recovery and it is not at all unique.
Early references concerning sufficiency in this quantum mechanical
setting are \cite{petz1988, petz1989} and our general reference is Chap. 
9 of \cite{OP}. (In classical mathematical statistics sufficiency is
a standard subject included in most books, our terminology is close
to \cite{Strasser}.)

Algebraicly $\beta$ in (\ref{E:M}) is the left inverse of $T$ as far as the 
densities $D_1$ and $D_2$ are concerned. It is easy to give an example 
where such a $\beta$ exists.  If $T$ is implemented by a unitary $U: \iH \to \iK$, 
then $\beta$ can be implemented by $U^* :\iK \to \iH$. This is a trivial situation. 
It is a bit less trivial that $\beta$ exists also in the case when $D_1=D_2$. 

The aim of this paper is to characterize the situation when the above $\beta$ 
exists. Actually, this was done a long time ago. It was proved in \cite{petz1988}
(see also \cite{mon2}) that $\beta$ exists if and only if
\begin{equation}\label{E:ns}
T^*\Big(T(D_2)^{\im t}T(D_1)^{-\im t}\Big)= D_2^{\im t} D_1^{-\im t} 
\end{equation}
for all real $t$, where $T^*$ is the standard transpose of $T$. Although this
is a necessary and sufficient condition, it is not completely satisfactory,
since it does not give any hint about the interrelation of $T$, $D_1$ and $D_2$.

The main result of the present paper is to show that (\ref{E:ns}) implies
the decomposition 
$$
D_s=\sum_{p=1}^r \lambda_s(p)S_s^\iH(p)R^\iH(p)\,,
$$ 
where  $S_s^\iH(p)$ commutes with $R^\iH(p)$, there are pairwise orthogonal 
projections $q_p$ such that $S_s^\iH(p)$ and $R^\iH(p)$ are supported in $q_p$ for
all $1 \leq p \leq r$ and  $\lambda_s(p)$ are some probability distributions 
($s=1,2$). The point is that the second factor is the same
for $s=1$ and for $s=2$.

Since the complete positivity of $T$ is not assumed, the Stinespring 
dilation cannot be used. For this reason and also due to the algebraic
methods, our approach is different from \cite{koashi}, where the
conditions $T(D_1)=D_1$ and $T(D_2)=D_2$ are studied and variety of physical
motivations is given.

We apply our structure theorem to deduce a sufficient and necessary 
condition for the equality case in the strong subadditivity of quantum
entropy and obtain the result of \cite{HJPW} as an application.

In the whole paper, an algebraic approach is followed.

\section{Preliminaries}
Let $\iH$ and $\iK$ be finite dimensional Hilbert spaces. Recall that
{\it 2-positivity} of $\tau: \BH \to \BK$ means that
$$
\left[ \begin{array}{cc}
\tau(A)&\tau(B)\\ \tau(C)&\tau(D)\end{array} \right] \geq 0 \hbox{\ if }
\left[ \begin{array}{cc}
A& B\\ C& D \end{array} \right] \geq 0\,. 
$$
It is well-known that a 2-positive unit-preserving mapping $\tau$ 
satisfies the {\it Schwarz inequality} $\tau(A^*A)\geq 
\tau(A)^*\tau(A)$.

The most important 2-positive mappings are of the form
\begin{equation}\label{E:cp}
\tau (A)=\sum_i L_iAL_i^*
\end{equation}
with some linear operators $L_i :\iH \to \iK$. (This is the 
{\it Kraus representation} of the completely positive maps.) We call
$L_i$ {operator coefficients}.
 
In this paper $T$ always denotes a trace-preserving 2-positive 
mapping $T : \BH \to \BK$ and we assume that the density matrices 
$D_1, D_2, T(D_1)$ and $T(D_2)$ are all invertible. If $T$ admits 
a Kraus representation, then the operator coefficients satisfy 
$\sum_i L_i^* L_i=I$. Lots of applications of such mappings are given in
\cite{N-Ch} in the setting of quantum information theory.

The spaces $\BH$ and $\BK$ are Hilbert spaces when they are endowed 
with the {\it standard Hilbert-Schmidt inner product} 
$$
\langle  A,B \rangle :=\tr A^*B\, .
$$
For a trace-preserving 2-positive mapping $T : \BH \to \BK$, 
its adjoint $T^*$ is a unital 2-positive mapping. 
It follows that $T^*: \BK \to \BH$ satisfies the Schwarz inequality.

The spaces $\BH$ and $\BK$ admit also the inner products
$$
\langle A,B \rangle_{D_1} := 
\tr A^*D_1^{1/2}BD_1^{1/2} \quad (A,B \in \BH )
$$
and
$$
\langle X,Y\rangle_{T(D_1)}: = \tr X^*T(D_1)^{1/2}YT(D_1)^{1/2} 
\quad (X,Y \in \BK ).
$$
The dual $\aa$ of $T^*$ with respect to these inner products is 2-positive
and unital, and it is characterized by the properties
\begin{equation}
\aa: \BH \to \BK, \quad \langle X, \aa(A)\rangle_{T(D_1)}=
\langle T^*(X), A \rangle_{D_1}\quad (A\in \BH,X\in \BK).
\end{equation} 
It is easy to give $\aa$ concretely:
\begin{equation}\label{E:alfa}
\aa(A)= T(D_1)^{-1/2} T\bz D_1^{1/2}A D_1^{1/2}\jz T(D_1)^{-1/2}
\end{equation}
It is seen from this formula that if $T$ has operator coefficients 
$L_i$, then the operator coefficients of $\aa$ are $T(D_1)^{-1/2}L_i D_1^{1/2}$. 

Note that if $T^*$ is an embedding, then $\alpha$ is the generalized conditional 
expectation introduced in \cite{AC}, see \cite{PD1984} for generalizations and for
a systematic study. This kind of dual was called transpose in \cite{OP} and makes
appearance in several places, for example in connection with the best quantum
recovery map \cite{BaK}, or in the theory of Connes-Narnhofer-Thirring dynamical
entropy \cite{OP}.

The standard dual $T^\#:\BK \to \BH$ of $\aa$ is trace preserving.
The next few lines follow simply from the definition of $T^\#$ and
the concrete form of the above inner products:
\begin{eqnarray*}
\tr T^\#(X) A &=& \langle T^\#(X^*), A \rangle =
\langle X^*, \aa(A) \rangle \\
&=&\tr X \aa(A)= 
\langle T(D_1)^{-1/2}X^* T(D_1)^{-1/2}, \aa(A) \rangle_{T(D_1)}\\
&=&\langle T^*(T(D_1)^{-1/2}X^* T(D_1)^{-1/2}), A \rangle_{D_1}\\
&=&\tr T^*(T(D_1)^{-1/2} X T(D_1)^{-1/2})D_1^{1/2} A D_1^{1/2}. 
\end{eqnarray*}
Hence
\begin{equation}\label{E:tehes}
T^\#(X)= D_1^{1/2} T^*\bz T(D_1)^{-1/2}XT(D_1)^{-1/2}\jz D_1^{1/2}.
\end{equation}
Observe that $T^\#(T(D_1))=D_1$. 

In the analysis of condition (\ref{E:M}) we first establish that the
existence of $\beta$ implies that from the set of all possible $\beta$'s
satisfying (\ref{E:M}) we can choose one canonically, namely $T^\#$. 
Remember that the definition of $T^\#$ depends on the density $D_1$, 
although this dependence is not included in the notation.

Assume now the existence of $\beta$ for (\ref{E:M}). According to 
Theorem 2 in \cite{mon2} we have (\ref{E:ns})
for all real $t$. Under our hypothesis $u_t:=T(D_1)^{\im t}T(D_2)^{-\im t}$ 
and $w_t:=D_1^{\im t} D_2^{-\im t}$ are unitaries and condition (\ref{E:ns}) 
tells us that $u_t \in \iA_{T^*}$ for every $t \in \bbbr$, see Lemma
\ref{L:2} below for $\iA_{T^*}$ and its properties. Consequently, 
$T^*(u_tY)=T^*(u_t)T^*(Y)$ and by analytic continuation we have
\begin{eqnarray*}
T^*\bz T(D_1)^{-1/2}T(D_2)T(D_1)^{-1/2}\jz &=&
T^*\bz T(D_1)^{-1/2}T(D_2)^{1/2}\jz 
T^*\bz T(D_2)^{1/2} T(D_1)^{-1/2}\jz
\\&=&D_1^{-1/2}D_2^{1/2} D_2^{1/2} D_1^{-1/2}\,
\end{eqnarray*}
which implies
\begin{equation}\label{E:4}
D_1^{1/2} T^*\bz T(D_1)^{-1/2}T(D_2)T(D_1)^{-1/2}\jz D_1^{1/2}=D_2\,.
\end{equation}
Therefore the relation $T^\#(T(D_2))=D_2$ can be concluded and in 
this way the following has been shown in \cite{petz1988}.

\begin{proposition}
If there exists a trace preserving 2-positive mapping $\beta : \BK \to 
\BH$ such that $\beta T (D_1)=D_1$ and $\beta T (D_2)=D_2$, then
$T^\# T (D_1)=D_1$ and $T^\# T(D_2)=D_2$.
\end{proposition}

Consider now the 2-positive unital mapping $(T^\# T)^*=T^*\circ \alpha
=:\gamma_\iH$. If $L_i$ are the operator coefficients of $T$, then
$\gamma_\iH$ has coefficients $L_j^*T(D_1)^{-1/2}L_i D_1^{1/2}$.
Let $\iD_\iH$ be the set of its fixed points. Since $\gamma_\iH$
leaves the states corresponding to $D_1$ and $D_2$ invariant, the
mean ergodic theorem applies and tells us the existence of conditional 
expectation $E$ from $\BH$ to $\iD_\iH$ which commutes with $\gamma_\iH$ 
and has the property $E^*(D_s)=D_s$ ($s=1,2$). Takesaki's theorem 
(\cite{Stra, Tak}, cf. Theorem 4.5 in \cite{OP}) tells us that
\begin{equation}\label{E:stab}
D_s^{\im t}\iD_\iH D_s^{-\im t}\subset \iD_\iH  \qquad (t \in \bbbr).
\end{equation}
(In another formulation, $\iD_\iH$ is stable under the modular groups, see 
Chapter 4 of \cite{OP} for a concise overview of the modular theory.) 

\section{Structure of certain unitaries}

In order to understand condition (\ref{E:stab}), we analyze the relation
$$
u^* \A u \subset \A
$$
for a unitary $u$ and for a subalgebra $\A$ of $\BH$. The result is  
formulated in the propositions below. We shall use the emerging 
structure in the next section but the result is interesting in itself.

Since $\A$ is finite dimensional, it is isomorphic to a direct
sum of full matrix algebras, so in an appropriate basis, elements 
of $\A$ have a blockdiagonal form
\begin{equation}
A=\oplus_{(m,d)} \bz \oplus_{i=1}^{K(m,d)} \bz 
\oplus_{t=1}^{m} A\bz m,d,i\jz\jz\jz,\label{E:A-forma} 
\end{equation}
where $m$ denotes the multiplicity and $d$ the dimension of the block $A(m,d,i)$.

For example, if there are three different blocks with multiplicity two, 
two of them with dimension two and one of them with dimension three, 
and another block with multiplicity four and dimension one, then 
$K(2,2)=2,\s K(2,3)=1,\s K(4,1)=1$  and every element $A\in\A$ has 
the form
\begin{align*}
A= \Diag (&B_1,B_1,B_2,B_2,C,C,d,d,d,d)\\
=\Diag(&A(2,2,1),A(2,2,1),A(2,2,2),A(2,2,2),A(2,3,1),
A(2,3,1),\\ &A(4,1,1),A(4,1,1),A(4,1,1),A(4,1,1)) 
\end{align*}
with $B_1,B_2\in\M_2$, $C\in\M_3$ and $d\in\M_1=\bbbc$.

Let $P_{m,d}$ be the projection in $\A$ corresponding to multiplicity 
$m$ and dimension $d$, that is
$$
P_{m,d}(m',d',i):=\delta(m',m)\delta(d',d)I_{d}\qquad (1 \le i \le K(m,d)),
$$ 
where $I_{d}$ is the $d \times d$ identity matrix, and let 
$$
P_m:=\oplus_{d} P_{m,d}
$$ 
be the projection corresponding to the multiplicity $m$. We denote by $\iH_m$
the range of $P_m$. Note that $P_{m,d}$ and hence $P_m$ commutes with 
elements of $\A$, so 
$$
\A_{m,d}:=P_{m,d}\A \ds\ds \mbox{ and }\ds\ds \A_m:=P_m\A
$$ 
are unital algebras with unit $P_{m,d}$ and $P_m$, respectively.

We fix an orthonormal basis 
$$
\{e(m,d,k)\s:\s 1\le k\le mdK(m,d)\}
$$ 
in the range of $P_{m,d}$ for every possible $m$ and $d$, such 
that $|e(m,d,k)\>\<e(m,d,k)|\in\A$.

\begin{proposition} \label{mult}
Let $u$ be a unitary in $\BH$ such that the map $A \mapsto u^*Au$ leaves 
$\A$ invariant. Then $u$ commutes with $P_m$ for every multiplicity $m$.
\end{proposition}

\proof
The statement is 
trivial when only one multiplicity exists, and we apply 
mathematical induction in the number of multiplicities.

Note that the rank of any minimal projection in $P_m \A$ is $m$. Let $m_1$ 
denote the smallest multiplicity, and let $q \in P_{m_1}\A$ be a 
minimal projection,
then $q$ is of rank $m_1$.  $P_m u^*qu$ is a projection again and its rank 
is at most $m_1$. Since all non-zero projections in $P_m \A$ has rank at 
least $m$, we can conclude that $P_m u^*qu=0$ if $m>m_1$. Every element 
of $P_{m_1}\A$ is a linear combination of the above $q$'s, hence we have 
$P_m u^*au=0$ if $m> m_1$ and $a \in P_{m_1}\A$.

Since $|e(m_1,d,k)\>\<e(m_1,d,k)|\le P_{m_1}$ for every possible 
$d,k$, it follows that 
$$
P_m u^* |e(m_1,d,k)\>\<e(m_1,d,k)| u P_m \le P_m\bz u^*P_{m_1}u \jz=0 
$$ 
if $m> m_1$. So we can conclude that 
$u^* e(m_1,d,k)\in \H_{m_1}$ which gives that both $\H_{m_1}$ and its 
orthogonal complement are invariant subspaces for $u$, that is, $P_{m_1}u=
uP_{m_1}$. Now we can restrict the whole problem to the orthogonal complement 
of $\H_{m_1}$ and use induction hypothesis in the number of multiplicities. 
\phantom {mmmm}\bv

We have obtained that $u$ has a blockdiagonal structure 
$u=\oplus_{m} P_m u P_m$, and to explore the finer structure of $u$, 
we can restrict our attention to the case when all the multiplicities 
are the same number $m$, i.e. the elements of $\A$ have the form
$$ 
A=\oplus_{d}\bz\oplus_{i=1}^{K(d)} \bz  
\oplus_{t=1}^{m} A\bz d,i\jz\jz\jz.
$$
As before, we can define projections $P_{d,i}$  ($1 \le i \le K(d)$)
by the formula
$$
P_{d,i}(d',i'):=\delta(d',d)\delta(i',i) I_d,
$$
and the projection corresponding to dimension $d$ is 
$$
P_d:=\oplus_{i=1}^{K(d)}P_{d,i}.
$$
Again, all these projections commute with all elements of $\A$.

\begin{proposition}
In the above setting $u$ commutes with $P_d$ for every dimension $d$, 
and so $u$ has the blockdiagonal structure
$$
u=\oplus_d P_d u P_d.
$$
\end{proposition}

\proof
Since $\A$ is isomorphic to $\M:=\oplus_{d} \oplus_{i=1}^{K(d)} \M_{d}$, 
then $\Ad_u$ induces an automorphism $\gamma$ of $\M$. The inclusion matrix 
of $\gamma$ is a permutation matrix corresponding to a permutation $\tau$ 
of the set of all possible pairs $(d,i)$ such that  
$$
\tau(d,i)=(d',i')\quad \mbox{gives}\quad d=d'.
$$
This implies that 
\[
P_{d'}u^*P_d u=0 \text{ for } d\ne d',
\]
which, by the same argument as in Proposition \ref{mult} implies the desired 
statement. \bv

In the view of the above Propositions we can suppose that all the blocks 
in $\A$ have the same multiplicity $m$ and the same dimension $d$, 
consequently $\A$ is isomorphic to $\oplus_{j=1}^K \M_d$. 
In this case $dim(\H)=mdK$, and $\B$ is isomorphic to $\M_d\otimes\M_m
\otimes\M_K$. Elements of $\A$ have the form
$$
\sum_i A_i\otimes I_m\otimes E_{ii},
$$
where $A_i$ is an element of $\M_d$ and $\{E_{ij} \s :\s 1 \le i,j\le K\}$ 
are the standard matrix units of $\M_K$.
It is easily seen that in this representation $u$ has the form
$$
u=\sum_{i=1}^K u_i\otimes E_{\sigma (i)i},
$$
where $\sigma$ is a permutation of the set $\{1,2,\dots ,K\}$, and 
the $u_i$ 's are easily seen to be unitary elements of $\M_d \otimes\M_m$ 
that leave the subalgebra $\M_d\otimes I_m$ invariant.
 
The final step to describe the structure of a possible unitary
$u\in\M_d\otimes\M_m$ such that $\Ad_u$ leaves the subalgebra $\M_d
\otimes I_m$ invariant. Since $\Ad_u$ induces an automorphism of $\M_d
\otimes I_m$, we have a unitary $v$ such that 
$$
\Ad_u(a\otimes I_m)=
(v\otimes I_m)^*(a\otimes I_m)(v\otimes I_m).
$$ 
The automorphism
$\Ad_u \circ \Ad_{v\otimes I_m}^{-1}$ leaves the subalgebra fixed and is
induced by a unitary $W\in \M_j\otimes\M_m$. Hence $W$ must be in the 
commutant of the subalgebra, that is, $W=I_d \otimes w$. From this we
conclude that $u=v\otimes w$.

We arrived at the following:

\begin{proposition}\label{P:3}
In the case when all the multiplicities and all the dimensions are the same,
$u$ must be of the form
$$
u=\sum_{i=1}^K v_i\otimes w_i\otimes E_{\sigma (i)i},
$$
where $v_i \in\M_d$ and $w_i \in \M_m$ are unitaries and 
$\sigma$ is a permutation of the set $\{1,2,\dots ,K\}$.
\end{proposition}
 
Note that $u \mapsto \sigma$ is a homomorphism on the group of 
allowed $u$'s (while $u \mapsto v_i\otimes w_i$ is not).
 
The general situation is put together from the above propositions:
$u$ commutes with $P_{m,d}$ and $uP_{m,d}$ is described by Proposition
\ref{P:3}.  

\section{Sufficient coarse-grainings}

Let $T : \BH \to \BK$ be a trace-preserving 2-positive mapping and
$D_t$ be density matrices on $\iH$ ($t=1,2$). We assume the existence
of a trace preserving 2-positive mapping $\beta : \BK \to \BH$ such that
$\beta \circ T (D_1)=D_1$ and $\beta \circ T (D_2)=D_2$. In other words,
we suppose that $T$ is sufficient for $D_1$ and $D_2$. Our goal is
to describe the structure coming from this assumption.

In this section we work with positive unital mappings, so are
the adjoint $T^* : \BK \to \BH$ and  $\aa:\BH \to \BK$ defined by
(\ref{E:alfa}). Let the fixed point algebra of $\gamma_\iH:= T^*\circ 
\aa$ be $\iD_\iH$,
that of $\gamma_\iK:=\aa\circ T^*$ be $\iD_\iK$. The mapping $\gamma_\iH$ 
leaves the state 
corresponding to $D_1$ invariant: $\Tr D_1 \gamma_\iH(A)=\Tr D_1A$ follows by 
easy computation. $\Tr D_2 \gamma_\iH(A)=\Tr D_2A$ was shown in the 
equivalent form $T^\#(T(D_2))=D_2$ in (\ref{E:4}). Similarly,
$\Tr T(D_1) \gamma_\iK(X)=\Tr T(D_1) X$ and $\Tr T(D_2) \gamma_\iK(X)=\Tr T(D_2)X$.

\begin{lemma}\label{L:2}
Let 
$$
\iA_{T^*}:=\big\{X\in \BK: T^*(XX^*)=T^*(X)T^*(X^*) \hbox{\ and\ }
T^*(X^*X)=T^*(X^*)T^*(X)\big\}
$$
and
$$
\iA_\aa:=\big\{A\in \BH: \aa(AA^*)=\aa(A)\aa(A^*) \hbox{\ and\ }
\aa(A^*A)=\aa(A^*)\aa(A)\big\}.
$$
Then $\iD_\iK \subset \iA_{T^*}$ and $\iD_\iH \subset \iA_{\aa}$. Moreover, 
$T^*$ restricted to $\iD_\iK$ is an algebraic isomomorphism onto 
$\iD_\iH$ with inverse $\aa$ and
\begin{equation}\label{E:alfaprop}
\aa(AB)=\aa(A)\aa(B)
\end{equation}
for all $A \in \iA_\aa$ and $B \in B(\iH)$.
\end{lemma}

The lemma is stated for reference, concerning the proof see 9.1 in
\cite{Stra}.

We have $T(D_1)^{\im t}\iD_\iK T(D_1)^{-\im t} \subset\iD_\iK $ and to 
the unitaries
$ T(D_1)^{\im t}$ we can apply the arguments in the previous section.  $A\in 
D_\iK $ has the form of (\ref{E:A-forma}) and we have the central projections
$\P(m,j)$ of  $\iD_\iK $ at our disposal. As above elements of $\P(m,j) \iD_\iK$
have the form
$$
\sum_i X_i\otimes I_m\otimes E_{ii},
$$
where $X_i$ is an element of $\M_j$ and $\{E_{ij} \s :\s 1 \leq i,j\leq k\}$ 
are the standard matrix units of $\M_k$. We can imagine $\P(m,j) B(\iK)\P(m,j) $ 
in the form $\M_j\otimes\M_m \otimes\M_k$.  

According to Proposition \ref{P:3}
every unitary $\P(m,j)T(D_1)^{\im t}$ is of the form
$$
\sum_{i=1}^k v_i\otimes w_i\otimes E_{\sigma (i)i},
$$
where $v_i \in\M_j$ and $w_i \in \M_m$ are unitaries and 
$\sigma$ is a permutation of the set $\{1,2,\dots ,k\}$, all of them 
depend on the real parameter $t$. Since this dependence is obviously 
continuous, the only possibility is $\sigma=$ identity. It follows that
$$
\P(m,j) T(D_1)=\sum_{i=1}^{k(m,d)} S_{11}(m,d,i)\otimes S_{12}(m,d,i)\otimes 
E_{ii}(m,d).
$$
Similar argument applies to $T(D_2)$ and we have
$$
\P(m,j) T(D_2)=\sum_{i=1}^{k(m,d)} S_{21}(m,d,i)\otimes S_{22}(m,d,i)
\otimes E_{ii}(m,d).
$$ 
If we want both factors to be normalized, then positive coefficients should 
be included in the front.

We refer to Theorem 9.11 from \cite{OP}, this tells that $T(D_1)^{\im t}
T(D_2)^{-\im t}$ belongs to $\iD_\iK$. Therefore 
\begin{equation}
S_{12}(m,d,i)=S_{22}(m,d,i) \qquad (1 \leq i \leq k(m,d))
\end{equation}

We want to see the densities $T(D_1)$ and $ T(D_2)$ in the central 
decomposition of the algebra $\iD_\iK$. Assume that $z_1,z_2,\dots,z_r$
the minimal central projections in  $\iD_\iK$. Then $z_p \iD_\iK$ is 
isomorphic to a full matrix algebra $\iM_{n_p}$ and  $\iD_\iK$ is 
isomorphic to $\oplus_{p=1}^r \iM_{n_p}$. In the above decomposition of
$T(D_1)$ we have
$$
S_{11} \otimes S_{12} \otimes E_{ii} = \big(S_{11} \otimes I\otimes E_{ii} 
\big)\big(I \otimes S_{12} \otimes E_{ii} \big),
$$
where the first factor belongs to a central summand $z_p \iD_\iK$ and
the second one is in $z_p \iD_\iK'$. 
Hence we arrived at the following structure.

\begin{thm}
Let $T : \BH \to \BK$ be a trace-preserving 2-positive mapping which is 
sufficient for the invertible density matrices $D_s$ on $\iH$ ($s=1,2$).
Assume that $T(D_1)$ and $T(D_2)$ are invertible as well. Then there
exists a subalgebra  $\iD_\iK \subset \BK$ with minimal central 
projections $z_1,z_2, \dots, z_r \in \iD_\iK$
such that
\begin{enumerate}
\item[(a)] $T^*(XY)=T^*(X)T^*(Y)$ for $X \in \BK$ and $ Y\in \iD_\iK$.
\item[(b)] $T(D_s)=\sum_{p=1}^r \lambda_s(p) S_s(p) R(p)$   
for some density operators $S_s(p) \in z_p \iD_\iK$ and $R(p)\in z_p 
\iD_\iK'$ and for probability distributions $\lambda_s(p)$ ($1 \leq p \leq r$)
for $s=1,2$.
\end{enumerate}
\end{thm}

The theorem is formulated in the Hilbert space $\iK$ but similar 
formulation is possible in $\iH$ as well. One starts with the observation
$T^*(\iD_\iK)=\iD_\iH \cong \oplus_{p=1}^r \iM_{n_p}$. $q_p:=T^*(z_p)$ are 
the minimal central projections in $\iD_\iH$. $S_s(p)=\aa(S_s^\iH(p))$ for
some $S_s^\iH(p) \in q_p \iD_\iH$ ($1 \le p \le r$). Property (\ref{E:alfaprop}) 
is reformulated for the standard dual $T^\#$ as
$$
A T^\#(B)=T^\#(\aa(A)B)
$$
and we have
$$
T^\#( S_s(p) R(p))=T^\#( \aa(S_s^\iH(p)) R(p))=S_s^\iH(p)T^\#( R(p)),
$$
therefore
$$
D_s=T^\#(T(D_s))=\sum_{p=1}^r \lambda_s(p)S_s^\iH(p)T^\#( R(p))\,
$$ 
where the support of $S_s^\iH(p)$ is in $q_p$ and $T^\#( R(p))$ commutes with
$S_s^\iH(p)$ for all $p$.

We first note that the structure formulated in the theorem is derived
from the sufficiency condition but on the other hand that structure 
implies sufficiency. Namely, the structure above guarantees condition
(\ref{E:ns}) by a simple calculation.
\begin{eqnarray*}
T^*\big(T(D_2)^{\im t}T(D_1)^{-\im t}\big)&=&T^*\Big(
\sum_{p=1}^r \lambda_2^{\im t}(p)\lambda_1^{-\im t}(p) 
S_2(p)^{\im t} S_1(p)^{-\im t}\Big)\\
&=& \sum_{p=1}^r \lambda_2^{\im t}(p)\lambda_1^{-\im t}(p) 
T^*\big(S_2(p)^{\im t} S_1(p)^{-\im t}\big)\\
&=& \sum_{p=1}^r \lambda_2^{\im t}(p)\lambda_1^{-\im t}(p) 
T^*\big(S_2(p)^{\im t}\big) T^*\big(S_1(p)^{-\im t}\big) \\
&=& \sum_{p=1}^r \lambda_2^{\im t}(p)\lambda_1^{-\im t}(p)
S_2^\iH(p)^{\im t} S_1^\iH(p)^{-\im t}\\
&=& D_2^{\im t}D_1^{-\im t}
\end{eqnarray*}

Our theorem extends the result in \cite{koashi} whose setting corresponds
to the case $\iH=\iK$ and $D_1=D_2$ in our notation and the decomposition
$$
\iH=\oplus_{p=1}^r  \iH_p^{left} \otimes \iH_p^{right}
$$
comes out.

Our theorem extends obviously to more density matrices. If $T$ is sufficient
for $D_1,D_2,\dots, D_k$, then all density matrices have the above form
and in each summand the first factor depends on $1\le s \le k$ while the 
second does not.
  
\section{Strong subadditivity of entropy}

The strong subadditivity of entropy is
$$
S(D_{ABC}) +S(D_B) \leq S(D_{AB})+S(D_{BC})
$$
for a system ${\cal H}_A\otimes{\cal H}_B \otimes {\cal H}_C$, where
$D_B, D_{AB}, D_{BC}$ are the reduced densities of the state $D_{ABS}$
of the composite system and $S$ stands for the von Neumann
entropy \cite{SSA}. We have the equivalent form 
$$
S(D_{ABC}, \tau_{ABC})+S(D_{B}, \tau_{B}) \geq  S(D_{AB}, \tau_{AB})
+S(D_{BC}, \tau_{BC})
$$
in terms of relative entropy \cite{OP}, $\tau$ denotes the density of the 
tracial state (for example, $\tau_{B}$ is $I_B/\dim {\cal H}_B$).   
This inequality is equivalent to the inequality
\begin{equation}\label{E:T}
S(D_{ABC},\tau _A \otimes D_{BC})\geq S(D_{AB}, \tau_A \otimes D_B).
\end{equation}
which, on the other hand, is the consequence of the monotonicity of
relative entropy. Uhlmann's theorem should be applied to the 
partial trace
$$
T(X \otimes Y \otimes Z)=  (X \otimes Y)\Tr Z\, ,
$$
since
$$
T(\tau _A \otimes D_{BC})=\tau_A \otimes D_B \quad \mbox{and} \quad
T(D_{ABC})=D_{AB}.
$$
To use our previous notation we set $\iH:={\cal H}_A\otimes
{\cal H}_B \otimes {\cal H}_C$, $\iK:={\cal H}_A\otimes
{\cal H}_B$, $D_1:=\tau _A \otimes D_{BC}$ and $D_2:=D_{ABC}$.
Our aim is to study the case of equality in (\ref{E:T}) which is
known to be equivalent of the sufficiency of $T$ with respect to
$D_1$ and $D_2$ (see \cite{petz1988} and \cite{mon2}).

We recall that $\iD_\iH$ is the fixed point algebra of the mapping
\begin{eqnarray*}
&&\gamma_\iH(X \otimes Y \otimes Z)=\\
&&\bz(\tau_A \otimes D_B)^{-1/2} T\bz (\tau _A \otimes D_{BC})^{1/2}
(X \otimes Y \otimes Z)
(\tau _A \otimes D_{BC})^{1/2} \jz (\tau_A \otimes D_B)^{-1/2}
\jz \otimes I_C\,.
\end{eqnarray*}
It is clear that $\gamma_\iH(X \otimes I_B \otimes I_C)=
X \otimes I_B \otimes I_C$, therefore 
$$
B(\iH_A)\otimes \bbbc I_B \otimes\bbbc I_C \subset \iD_\iH
\subset B(\iH_A)\otimes B(\iH_B)\otimes\bbbc I_C
$$
and $\iD_\iH$ must be of the form $B(\iH_A)\otimes \iA_B 
\otimes\bbbc I_C$ with a subalgebra $\iA_B$ of $B(\iH_B)$.
Elements of $\iA_B$ have the form (\ref{E:A-forma}) and
$$
I_B=\sum_{m,d}\P(m,j)\,,
$$
where the (central) projection $\P(m,j)$ corresponds to multiplicity $m$
and dimension $d$.

The algebra $\P(m,j) \iA_B \P(m,j)$ is isomorphic to $\oplus_{t=1}^{K(m,d)} 
\M_d$. In this case $\dim \P(m,j)=mdK$ and elements of $\P(m,j) \iA_B \P(m,j)$ 
have the form
$$
\sum_i B_i\otimes  E_{ii} \otimes I_m\, ,
$$
where $B_i$ is an element of $\M_d$, $\{E_{ij} \s :\s 1 \leq i,j
\leq K(m,d)\}$ are the standard matrix units of $\M_{K(m,d)}$
and $I_m \in \iM_m$ is the identity. We use these facts to pass to the
algebra $\iD_\iH$.  $P'_{m,d}:=I_A \otimes  P_{m,d} \otimes I_C$ is a
central projection and elements of $\iD_\iH P'_{m,d}$ are of the form
$$
\sum_i A \otimes B_i\otimes  E_{ii} \otimes I_m\otimes I_C.
$$

Since the unitaries $D_2^{\im t}$ commute with
$ P'_{m,d}$ we have
$$
D_2^{\im t}\bz \iD_\iH  P'_{m,d}\jz D_2^{-\im t} \subset\iD_\iH  P'_{m,d}
$$
and this allows us to establish the structure of $D_2  P'(m,j)$.
\begin{equation}
D_2  P'_{m,d}=\sum_i D_{AB1}(i)\otimes  E_{ii} \otimes D_{B2C}(i)\,,
\end{equation}
where $D_{AB1}(i)$ and $ D_{B2C}(i)$ are density matrices in $B(\iH_A)
\otimes \iM_j$ and $\iM_m \otimes B(\iH_C)$, respectively. We can conclude
the form of $D_{ABC}$ which allows equality in the strong subadditivity
for the entropy:
\begin{equation}\label{E:SSA}
D_{ABC}=\sum_{m,d}\sum_i^{K(m,d)} \lambda(i,m,d)D_{AB1}(i,m,d)\otimes  
E_{ii}(m,d) \otimes D_{B2C}(i,m,d)\,,
\end{equation}
where $I \otimes E_{ii}(m,d)\otimes I$ is pairwise orthogonal family of 
projections acting
on $\iH_B$. This structure is the same as the one obtained in \cite{HJPW}.

It has been known for a while that the equality in several strong subaddtivity
inequalities for the von Neumann entropy of the local restriction of states
of infinite product chains is equivalent to the Markov property initiated
by Accardi (see Proposition 11.5 in \cite{OP} or \cite{petz1994}). 
Therefore, from the
structure (\ref{E:SSA}), one can deduce the form of quantum Markov states
which was done in \cite{AL, Ohno} by different methods, see these papers
concerning the details.

\end{document}